\documentclass[letterpaper,aps,prb,floatfix,twocolumn,showpacs,superscriptaddress,10pt]{revtex4-1}
\usepackage{graphicx,times}
\usepackage{amssymb,amsmath}
\usepackage{multirow}
\usepackage{color}
\usepackage{epstopdf}
\usepackage{bm}
\usepackage{longtable}
\usepackage{lipsum}

\def\be{\begin{equation}}
\def\ee{\end{equation}}
\def\ba{\begin{eqnarray}}
\def\ea{\end{eqnarray}}

\begin{document}

\title{{Noise Assisted  Excitation Energy Transfer in a Linear Toy Model of a Selectivity Filter Backbone Strand}}

\author{Hassan Bassereh}
\affiliation{Department of Physics, Isfahan University of Technology, Isfahan 84156-83111, Iran}

\author{Vahid Salari}
\affiliation{Department of Physics, Isfahan University of Technology, Isfahan 84156-83111, Iran}

\author{Farhad Shahbazi}
\affiliation{Department of Physics, Isfahan University of Technology, Isfahan 84156-83111, Iran}

\date{\today}

\begin{abstract}
In this paper, we investigate the effect of noise { and disorder} on the efficiency of excitation energy transfer (EET) in a $N=5$ sites linear chain with "static" dipole-dipole couplings. In fact, {here}, the disordered chain is a toy model for one strand of the selectivity filter backbone in ion channels. { It is recently discussed that the presence of quantum coherence in the selectivity filter is possible  and  can play a role in mediating ion-conduction and ion-selectivity in the selectivity filter}. The question is "how a quantum coherence can be effective in such structures while the environment of the channel is dephasing (i.e. noisy)?"  
Basically, we expect that the presence of the noise  should have a destructive effect in the quantum transport. In fact, we show that such expectation is valid for ordered chains. { However,  our results indicate that introducing the dephasing in the disordered chains leads to the weakening  of the localization effects, arising from  the randomness,  and then increases the efficiency of quantum energy transfer.} { Thus,  the presence of noise is crucial  for the enhancement of EET efficiency in  disordered chains. 
 We also show that the contribution of both classical and quantum mechanical effects are required to improve the speed of  energy transfer along the chain.
Our analysis may help for better understanding of fast and efficient functioning of the selectivity filters in ion channels}
\end{abstract}


\maketitle

\section{Introduction}

Energy or charge transfer is one of the most important phenomena in physical and biological systems. Life-enabling transport phenomena in the molecular mechanism of biological systems occur at scales ranging from atoms to large macro-molecular structures. Charge transfer through DNA~\cite{1} or charge and  energy transfer processes in photosynthetic structures~\cite{2, 3, 4,5} are good examples in this context. Recently, it has been put forward the idea  that quantum mechanics  might have positive effect on the efficiency of energy or charge transfer in living systems.
One of the most important effects of quantum mechanics in biological systems has been evidenced in the {Fenna-Matthews-Olson (FMO)} complexes~\cite{6} which is observed by experimental methods via ultrafast spectroscopy~\cite{7}. 

\begin{figure}[t]
   \includegraphics[scale=0.25]{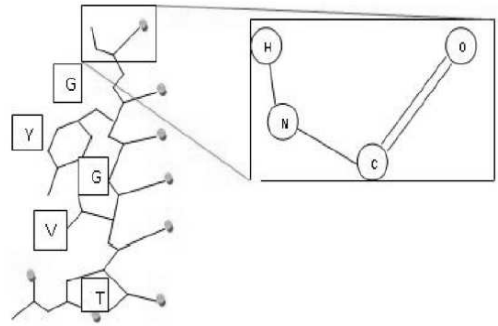}
       \caption{ P-loop strand of a selectivity filter: the amino acid molecular chains play the main role in conducting the ions. In the structure of the selectivity filter N-C=O is an amide group and C=O is a carbonyl group. Here we have considered each peptide unit H-N-C=O as one molecule. The P-loop strand is in the form of a chain of peptide units.}
            \label{fig1}
\end{figure}

\begin{figure}
    \includegraphics[scale=0.25]{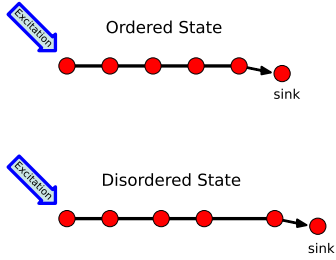}
        \caption{(Color online) Schematic representation of excitation energy transfer through (top) an ordered chain in which the distances between sites are equal and  (bottom) a disordered  chain with random distances between the molecules. }
             \label{fig2}
\end{figure}

Natural systems, inevitably, suffer from various types of noise with internal and external sources, which perturb the intrinsic dynamics of the real systems.    
Noise exists everywhere especially in the hot, wet and complex structures of biological systems. As a general expectation, noise is a destructive factor for any system, but in some living systems it is vice versa~\cite{8}.
Here, we would like to study excitation energy transfer in a toy model of a strand of a protein (i.e. a linear harmonic chain) in presence of noise via a quantum mechanical approach. It is known that in many ion-channel proteins, flow of ions through the pore is governed by a gate, comprised by a so-called selectivity filter that can be activated by electrical, chemical, light, thermal and/or mechanical interactions. The selectivity filter is believed to be responsible for the selection and fast conduction of particular ions across the membrane of an excitable cell. A selectivity filter is composed of four P-loop chains { (P-loop or phosphate-binding loop is a motif in proteins that is associated with phosphate binding)}, and each strand is composed of the sequences of TGVYG amino acids [T(Threonine, Thr75), V(Valine, Val76), G(Glycine, Gly77), Y(Tyrosine, Tyr78), G(Glycine, Gly79)] linked by peptide units H-N-C=O (see Fig.~\ref{fig1}). We consider a simplified model for one of the above chains in the selectivity filter in absence of ions, in which the system is made up of $5$ sites. Each site, being  equivalent to a peptide unit linked to an amino acid, is considered as an electric dipole, so the interaction energy between each two sites is assumed to be  dipole-dipole type following the inverse cube law~\cite{9}. Asadian et al ~\cite{9} considered a rather abstract system, a linear chain of $N$ molecules, and studied the effect of mechanical vibrations of the system on the  energy transfer across the chain.  They demonstrated that the collaborative interplay between the quantum-coherent excitation and the mechanical motion of the molecules would enhance the excitation energy transfer through the linear chain. The general issue is that an excitation is injected to the first site and after time $t$, the transferred energy to an additional site (i.e. sink) is measured ( e.g. sink can be located after G79 (or G) in Fig.~\ref{fig1}). A schematic representation of this energy transfer is illustrated in Fig.~\ref{fig2}. We also investigate the effect of  noise and disorder on  the energy transfer through the structure.

 Paper is organised as the following.  In Sec.~\ref{model} the model is described and quantum transport equations in terms of Lindblad operators are introduced. Excitation energy transport calculations, in absence and presence of noise, are given in Sec.~\ref{static}.  In Sec.~\ref{Hybrid}, we present an alternative approach to find  the contribution  of classical dynamics in the efficiency of energy transfer. The results are summarised in Sec.~\ref{conclusion}. 

\section{Structure and Transport Model}
\label{model}

Since there are five main peptide units in the P-loop structure, we take $N = 5$.  However, we are interested in measuring the amount of energy transported to the end of the P-loop strand.  As a result, an extra site so-called 'sink' will be introduced at the end of the loop \cite{8}. So, we consider a linear chain of $N=5$ sites, composed of two level molecules.  Each molecule can be stimulated to its excited state and the  excitation can be transferred to the other sites due to dipole-dipole couplings between the molecules. For simplicity, we consider only the  interaction between  nearest neighbour sites. Thus, the Hamiltonian of the system for a single excitation can be written in tight-binding form~\cite{9}
\begin {equation} 
H=\sum_{n=1}^{N} \varepsilon  |n\rangle\langle n|+ \sum_{n=1}^{N-1}J_n(|n\rangle\langle n+1|+|n+1\rangle\langle n|),
\label{Hamiltonian}
\end {equation}
where $|n\rangle$ denotes the excited state in the $n$-th site, and $\varepsilon$ is the excitation energy for a single molecule. 
 The dipole-dipole coupling between the sites $n$ and $n+1$ has the form~\cite{9}
\begin {equation} 
J_n=\frac{\tilde{J_0}}{{d_n}^3}, 
\label{coupling}
\end {equation}
in which $\tilde{J_0}$ contains the dipole  moment and is set to unity ( $\tilde{J_0}=1$), ${d_{0:n,n+1}}$ is the equilibrium distance between two neighbourings.
Since we consider the excitation in an open quantum system,  therefore  energy dissipation due to interaction with the environment is inevitable. Markov approximation formulation of  the  energy loss  due to dissipation, can be done by using the following Lindblad super-operator ~\cite{8}:
\begin {equation} 
L_{\rm diss}\rho=\sum_{n=1}^{N}\gamma_n[2{\sigma}^-_n\rho{{\sigma}^+_n}-\big\{{{\sigma}^+_n{\sigma}^-_n,\rho}\big\}],
\label{dissipation}
\end {equation}
where { $\rho$ is the total density matrix of the system}, ${\sigma}^+({\sigma}^-)$ is the creation (annihilation) operator of the excitation at site $n$, $\gamma_n$ is the dissipation rate at each site, and $\big\{{A,B}\big\}=AB+BA$.  
In order to measure how much of the excitation energy is transferred along the chain (and not lost due to dissipation), we introduce an additional site, the \emph{sink}, representing the final ($N+1$)-th trapping site that resembles a reaction center, in a photosynthesis system~\cite{8}. The sink is populated via irreversible decay of excitation from the last site, $N$. This approach is formally implemented by adding the following Lindblad operator to the master equation~\cite{8}:
\begin {equation} 
 L_{\rm sink}\rho=\gamma_{\rm {sink}}[2{\sigma}^+_{N+1}{\sigma}^-_N\rho{\sigma}^+_N{\sigma}^-_{N+1} -\big\{{{{\sigma}^+_N{\sigma}^-_{N+1}\sigma}^+_{N+1}{\sigma}^-_N,\rho}\big\}],
 \label{sink}
\end {equation}
{ in which  $\gamma_{\rm {sink}}$ denotes the absorption rate of the sink, for which we select the typical values $\gamma_{\rm {sink}}=0.1$ in this work. Later, we will discuss the dependence of the results on $\gamma_{\rm {sink}}$.}

In order to calculate the efficiency of energy transfer given by the asymptotic population of the sink, it is necessary to integrate the following master equation (${\hbar}=1$):
{
\begin {equation} 
\frac{\partial\rho}{\partial{t}}=i[\rho ,H]+L_{\rm diss}\rho+L_{\rm sink}\rho.
\label{emo}
\end {equation}
}

Here, an excitation is injected at the first site, i. e.  $\rho(0)=|1\rangle\langle 1|$ and we are interested in finding how much of the excitation can be transferred to the sink, so we have to find $P_{\rm sink}(t)=\langle N+1|\rho(t)|N+1\rangle$, which is called \emph{Sink population} (at time $t$). In this paper, we  use the python package  QUTIP ~\cite{11} to numerically solve the Lindblad master equations. In our numerical simulations, all energies, time-scales, and rates will be expressed in units of $\varepsilon$, and thereby we consider $\varepsilon = 1$.

\section{EET in Ordered and Disordered Chains}
\label{static}

\begin{figure}[ht]
\includegraphics[scale=0.33]{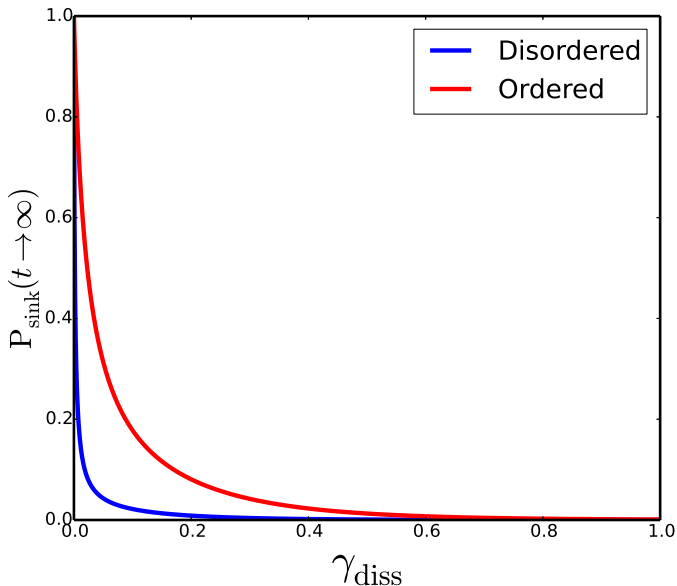}
\caption{ { (Color online)  The effect of dissipation rate, $\gamma_{\rm diss}$, on the system efficiency for ordered (blue (dark) line) and disordered (red (grey) line) chains  for $\gamma_{\rm {sink}}=0.1$.}} 
\label{p-diss}
\end{figure}   

In this section we study the quantum excitation energy transfer in the static chains in which the distances between sites are time independent. The ordered static chain is a chain in which the distances between sites are equal and the couplings between sites are frequency-independent and don't vary with time. A disordered static chain is the same as ordered static chain but the distances between sites are chosen  randomly (see Fig.~\ref{fig2}). 
 
The coupling between neighbouring sites is expressed as the following
 
\begin {equation} J_n=\frac{\tilde{J_0}}{[{d_{0:n,n+1}}]^3},
\end {equation}
Where $\tilde{J_0}$ is the dipole moment and $d_{0:n,n+1}$ is the equilibrium dimensionless distance between the sites $n$ and $n+1$, which is taken to be unity for ordered chain. For the disordered chain, we use the relative ratios of distances based on the real data obtained from the distances between amino acids in one strand of the backbone of the selectivity filter of KcsA ion channels. In the mentioned structure, the distances between neighbouring sites are, $r_{0:1,2}=2.6A^{\circ}, r_{0:2,3}=3.1A^{\circ}, r_{0:3,4}=2.6A^{\circ}$ and  $r_{0:4,5}=4.3A ^{\circ}$ ~\cite{10}, therefore dividing the distances over the minimum one $2.6A^{\circ}$, gives us the relative distances  as $d_{0:1,2}=1,d_{0:2,3}=1.19, d_{0:3,4}=1$ and $d_{0:4,5}=1.65 $. Energy population in the sink is calculated using the method explained in the Sec.~\ref{model}. 
{ As a matter of fact, each molecule feels a different environment, hence the local dissipation rates  $\gamma_n$ depend on the sites, nevertheless for the sake of simplicity and due to so far unavailable experimental data,  we will later assume them to be uniform $\gamma_{n}=\gamma_{\rm {diss}}$. In order to pick a typical value for dissipation rate, we fixed $\gamma_{\rm {sink}}=0.1$ in Eq. \ref {emo} and calculated the sink population in terms of  $\gamma_{\rm diss}$. The results  represented  in Fig.~\ref{p-diss}, show the rapid decaying of sink population with  dissipation rate coefficient in both disordered and ordered static chains.  Since we are interested in system with efficient energy transfer, hence in the rest of the paper we  select a small value $\gamma_{\rm diss}=0.001$ for the dissipation rate.}

\begin{figure}[t]
    \includegraphics[scale=0.3,width=8cm]{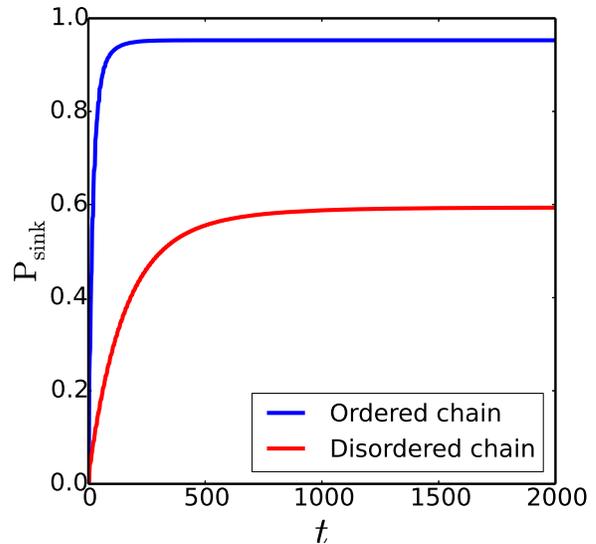}
        \caption{ {(Color online)Temporal variation of sink population for static ordered (blue (black) solid line) and disordered chain (red (grey) solid line). } } 
        \label{p-t}
\end{figure}

\begin{figure*}[t]
    \includegraphics[scale=0.4]{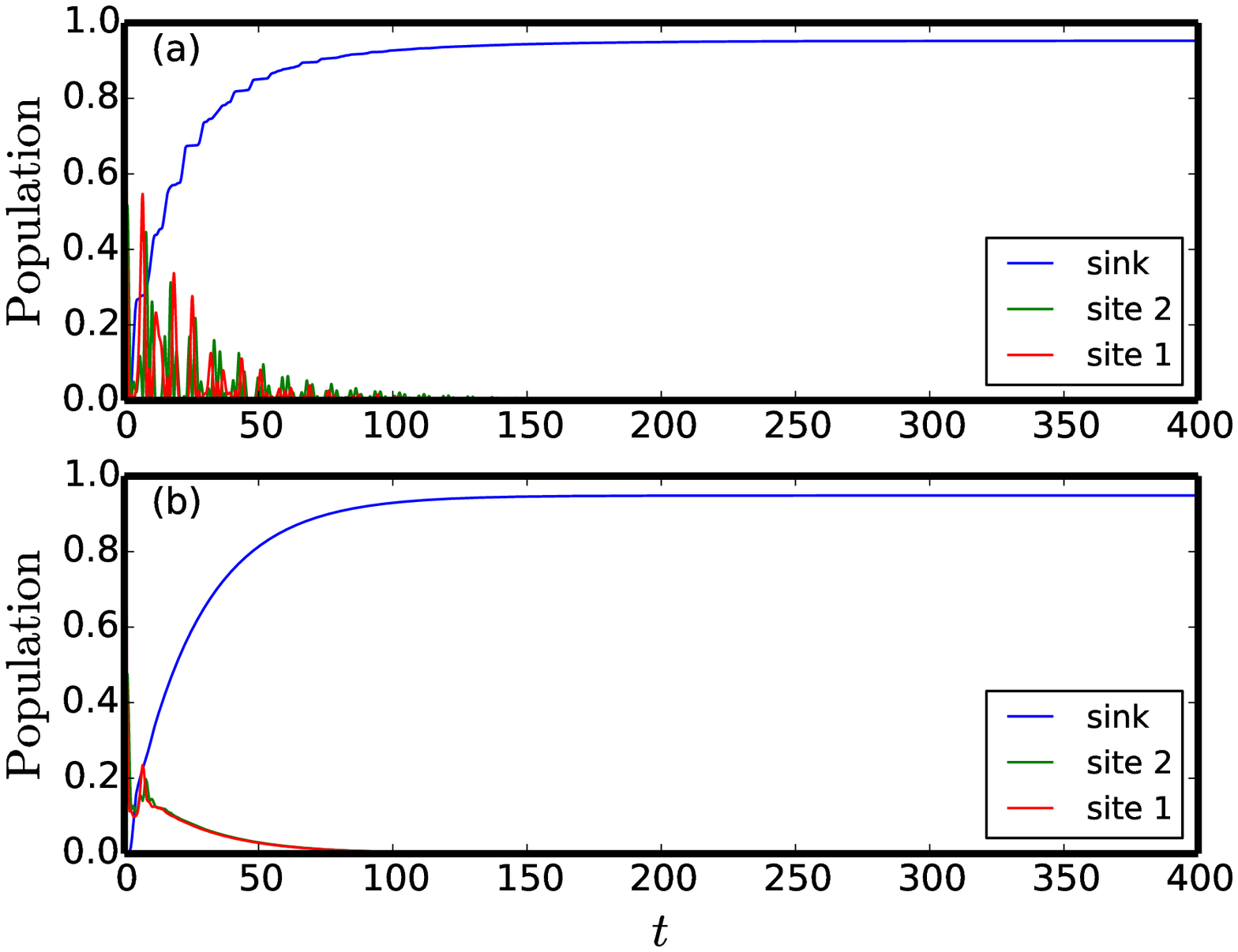}
    \includegraphics[scale=0.4]{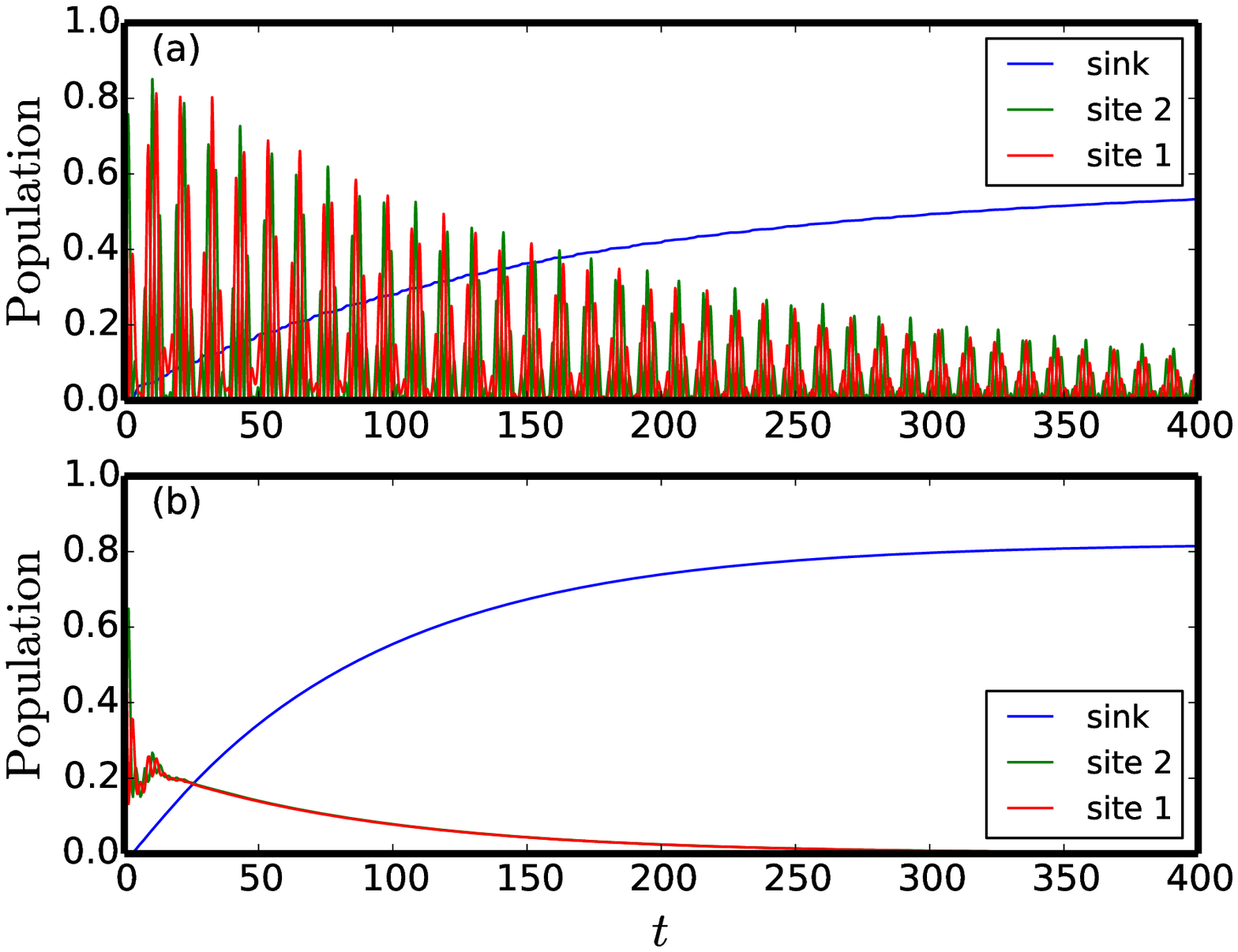}
        \caption{ {(Color online) Left) Temporal variation of  energy population of the sites $1$, $2$ and the sink in  ordered chain  for (a) absence and (b) presence of dephasing with $\gamma_{\rm depth}=0.2$. Right) Temporal variation of  energy population of the sites $1$, $2$ and the sink in disordered chain  for (a) absence and (b) presence of dephasing with $\gamma_{\rm depth}=0.2$} } 
        \label{loc}
\end{figure*}

{ With such a selection for dissipation, the sink population are plotted in Fig.~\ref{p-t}versus time for ordered and disordered chains. It can be seen that population for disordered chain is clearly less than the disordered one, which is a result of 
the Anderson localization~\cite{localization} of  the eigenstates of the tight-binding Hamiltonian in a one dimensional disordered chain. In fact, the randomness in the disordered chain plays the role of scattering centres for the Bloch states corresponding to the ordered chain. The constructive interference, resulting from successive coherent back scatterings of the Bloch states by such scattering centres,  would lead to the localization of these states within  a region characterised by the localization length. Decomposing  the initial excitation into the eigenstates of the original Hamiltonian \ref{Hamiltonian},  the modes with higher energy have smaller localization length and are more probable to become  localized.  The localised modes spend much time inside the chain and then leak out to the environment because of dissipation, hence loss their ability to corporate in the  energy transfer into the sink. To explicitly show the localization effect on the energy transfer, we plot the time dependence energy density in the first two sites  in the left-(a) and right-(a) panels of  Fig. \ref{loc}  for ordered and disordered chains, respectively. The temporal behaviour of energy populations in the other sites are  similar to the the first two sites and for the sake of clarification we do not show them. It can be clearly seen that the energy population in the sites $1$ and $2$ persists much longer  in the disordered chain than the ones in the ordered chain.}

 \begin{figure}[h]
        \includegraphics[scale=0.32]{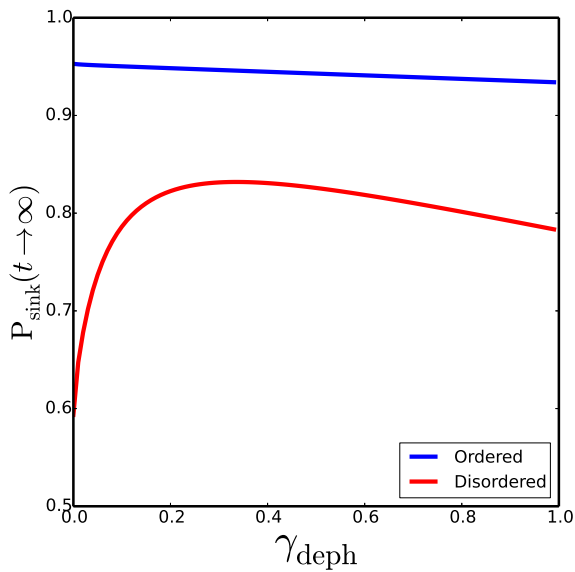}
        \caption{ {(Color online)  Long time sink population of static ordered (blue (black) solid line) and disordered (red (grey) solid line) chains versus dephasing rate coefficient.} } 
        \label{p-deph}
\end{figure}

\begin{figure*}[t]
\includegraphics[scale=0.35]{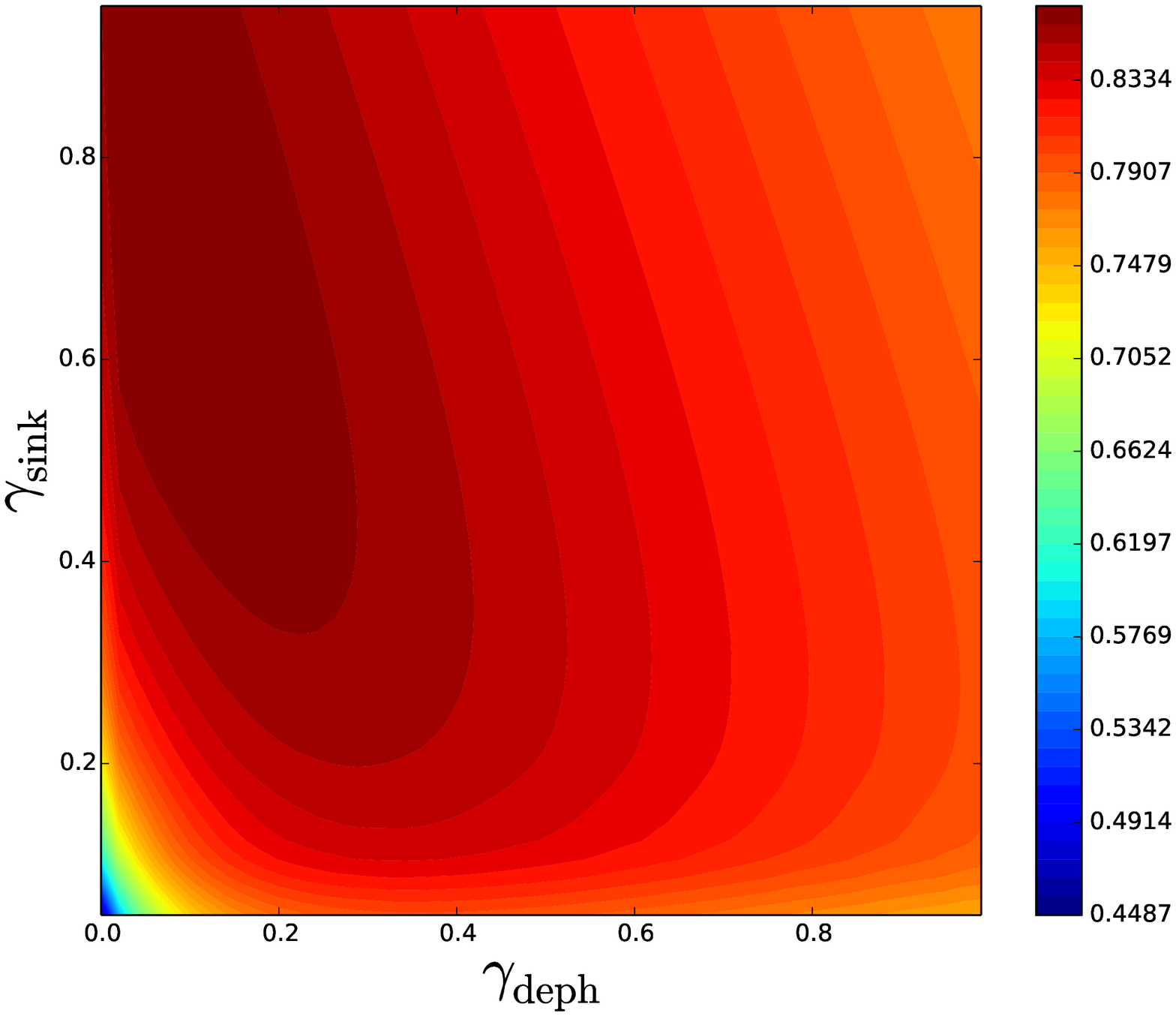}
\includegraphics[scale=0.35]{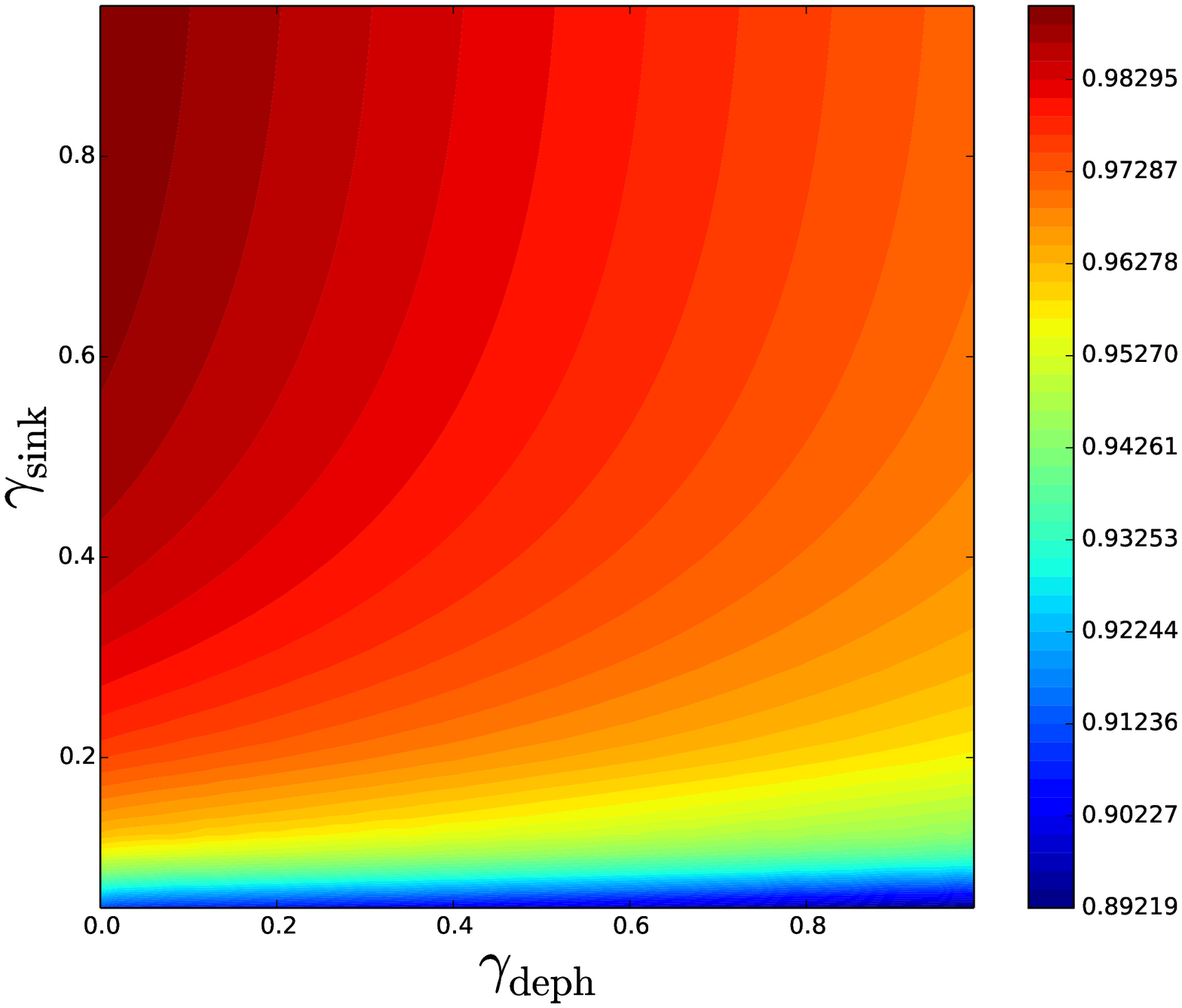}
\caption{ { (Color online) The density plots of sink population  for the disordered  chain (left) and ordered  chain (right) with $\gamma_{\rm{diss}}=0.001$.}} 
\label{density}
\end{figure*}

{Now we proceed to investigate the effect of noise on the quantum energy transfer along the chain}. 
Real systems are open and  interact with environment, so it  is reasonable to assume that the environment is dephasing which means it is noisy and has a destructive effect on quantum coherence in the system. The dephasing term can   be incorporated to  the master equation, using   the following Lindblad form~\cite{8}
\begin {equation}
 L_{\rm deph}\rho=\sum_{n=1}^{N}\gamma_{\rm deph}[2{\sigma}^+_n{\sigma}^-_n\rho{{\sigma}^+_n{\sigma}^-_n}-\big\{{{\sigma}^+_n{\sigma}^-_n,\rho}\big\}]
 \label{dephase}
 \end {equation}
where $\gamma_{\rm deph}$ is the dephasing rate coefficient.
In general, one expects decreasing population after adding dephasing term. 

{Fig.~\ref{p-deph}} illustrates  variation  of the sink population, as $t\to\infty$, versus dephasing rate for the ordered and disordered chains. As we expected, dephasing (i.e. noise) decreases population (i.e. efficiency) of the ordered chain. However, something different is observed for the static disordered chain, it is seen that the sink population does not behave monotonically versus dephasing rate, in such a way that  noise is able to increase the efficiency of energy transfer  and reach it to a maximum efficiency for $\gamma_{\rm {depth}}\sim 0.2$. {The reason for  noise-assisted energy transfer in the disordered chain is the weakening of the localization,  as a result of dephasing  which reduces the  constructive coherent back-scatterings responsible for such a  localization, hence leading to the enhancement of the efficiency of  energy transfer. The panels (b) in Fig. \ref{loc} evidently represent the demolishing effect of dephasing on the constructive back scattering in both ordered and disordered chains. As can be seen from the left-(b) panel of Fig~\ref{loc}, the destruction of Anderson localization in the disordered chain leads to  remarkable  enhancement in the amount of the sink population, and  also the speed of energy transfer. }

{ In order to gain insight into  the effect of possible values of the absorption rates of the sink, $\gamma_{\rm {sink}}$, and dephasing rates, $\gamma_{\rm {deph}}$ on the energy transfer efficiency, we illustrate    the density plots of  sink population for the ordered (Fig.~\ref{density}, left) and disordered (Fig.~\ref{density}, right) chains. This figure shows there is no region in the parameter space, $\gamma_{\rm {sink}}-\gamma_{\rm {deph}}$, for the noise-assisted energy transfer in the ordered chain, while  we see that such a phenomenon occurs in the disordered chain in the region with $ \gamma_{\rm {sink}}\lesssim 0.5$.}

\section{hybrid quantum-classical dynamics} 
\label{Hybrid}
 { We have demonstrated that the collaborative interplay between the quantum-coherent excitation and the mechanical motion of the molecules would enhance the excitation energy transfer through the linear chain. It is already concluded that a closer look at the involved dimensions and energetics  of the process reveals that the underlying mechanism for ion transmission and selectivity might be not entirely classical \cite{12}. Recently it was demonstrated theoretically that, based on the time and energetic scales involved in the selectivity filter, the ion selectivity and transport cannot be entirely a classical process but involves quantum coherence \cite{13}.} { In this section we pursue an alternative approach to evaluate  the  contribution of the classical and quantum mechanical effects in the efficiency of energy transfer across  the chain.  For this purpose, we use  Kossakowski-Lindblad master equation \cite{Caruso2014, Kossakowski} in the following form:   
       
 \begin{equation} 
\frac{\partial\rho}{\partial{t}}=-(1-\eta)i[H,\rho]+\eta \sum_{ij}[L_{ij}\rho L_{ij}^\dag -1/2\big\{{L_{ij}^\dag L_{ij},\rho}\big\}],
\label{emo2}
\end{equation}
where $\eta$ is a measure for  the classical contribution of the energy transfer. If $\eta=0$ we have a pure quantum energy transfer and if $\eta=1$ the energy transfer is fully calssical \cite{Caruso2014}, and for the other values of $\eta$ (between 0 and 1) we have a joint cooperation between classical and quantum mechanical effects. In the above equation, $L_{ij}=T_{ij}|i\rangle\langle j|$, where $|i\rangle$ denotes the real space basis state  and we select $T_{ij}=J_0/d_{ij}^3$, in which $J_{0}=1$ and $d_{ij}$ is selected as the real normalized distance between the sites i and j( amino acids in the p-loop strand) in the disordered chain and $d_{ij}=1$ for the ordered chain.

\begin{figure}[b]
\includegraphics[scale=0.4]{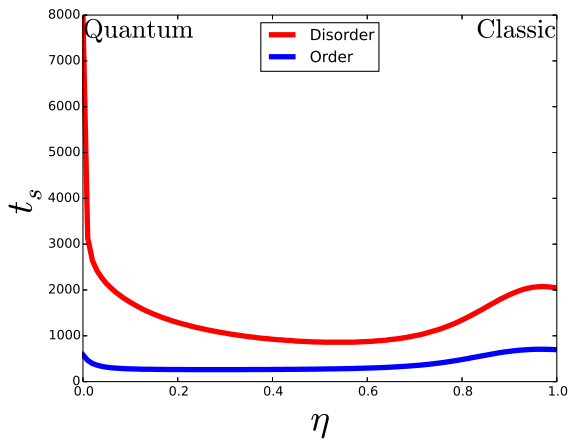}
\caption{ (Color online)  Saturation time versus the coefficient of  classical part of energy transfer dynamics }
\label{hybrid}
\end{figure}

Here we neglect the dissipation and again add $L_{\rm {sink}}$ to the above Eq. \ref{emo2} and calculate the saturation  time $t_s$ ( the time which all of the first injected energy is transferred  to the sink) versus parameter $\eta$. As  shown in Fig.~\ref{hybrid}, $t_s$ is not much sensitive to $\eta$ in the ordered chain, however in the disordered chain we observe that $t_s$  rapidly drops by increasing the classical part of the energy transfer dynamics.  The optimal stationary time is obtained for $\eta\sim 0.5$, meaning that the maximum efficiency in the speed of energy transfer is acquired by approximately equal contribution of classical and quantum mechanical  effects. 
}

 \section{Discussion and Conclusion}
 \label{conclusion} 

 { Ion channels are proteins in the membranes of excitable cells that cooperate for the onset and 
propagation of electrical signals across membranes by providing a highly selective 
conduction of charges bound to ions through a channel like structure. The selectivity filter is a part of the 
protein forming a narrow tunnel inside the ion channel which is responsible for the selection 
process and fast conduction of ions across the membrane \cite{Summhammer2012}.
 Certainly, the magnitude of the thermal fluctuations of the backbone atoms forming the selectivity filter is large relative to the small size difference between Na and K, raising fundamental questions about the mechanism that gives rise to ion selectivity. This suggests that the traditional explanation of ionic selectivity should be reexamined \cite{Roux2004}.}
 Recently, Vaziri et al.~\cite{12} and Ganim et al.~\cite{13} proposed the presence of quantum coherence in $K^+$ ion-channels,  in their backbone amide groups that can play a role in mediating ion-conduction and ion-selectivity in the selectivity filter. 
  In summary, we analysed quantum excitation energy transfer (EET) in a linear  chain composed of five sites, as a toy model for one strand of selectivity filter backbone in ion channels. The inter-site separations  is adjusted to be proportional to the distances  between peptide units in the selectivity filter. EET is studied both   in absence and presence of dephasing noise. Comparison of the result with those obtained in an ordered chain, indicates that disorder in such systems has  always a destructive role in EET, because of  the Anderson localization effect. 
When dephasing is introduced, it is found that  noise has  destructive effect in EET in ordered chains. Nevertheless, for disordered chains  it is shown that the noise is able to significantly  increase the efficiency of energy transfer across the chain either in amount and speed. 
{ The main message of this work is  the significance of dephasing in the efficiency of quantum energy transfer in disordered chains}. The  living systems are mostly disordered and in some cases are very efficient, so it is possible in such systems that noise has a key role for efficient energy transfer through their structures, e.g. selectivity filter backbone.

\end{document}